\documentstyle[aps,pre,multicol,epsf]{revtex}
\def\be{\begin{equation}}
\def\ee{\end{equation}}
\def\bea{\begin{eqnarray}}
\def\eea{\end{eqnarray}}

\newcommand{\f}{\varphi}
\newcommand{\p}{\partial}
\newcommand{\e}{\varepsilon}

\begin{document}
\draft

\title{Concentration Profiles and Reaction Fronts in 
A$+$B$\rightarrow$C Type Processes:\\
Effect of Background Ions}

\author {T. Unger and Z. R\'acz}
\address{Institute for Theoretical Physics,
E\"otv\"os University,
1117 Budapest, P\'azm\'any s\'et\'any 1/A, Hungary}

\date{\today}

\maketitle

\begin{abstract}
Diffusion and reaction of initially separated ions 
$A^-$ and $B^+$ in the presence of counter ions ${\hat A}^+$ and 
${\hat B}^-$ is studied. The dynamics is described in terms of 
reaction-diffusion equations obeying 
local electroneutrality, and the time-evolution of
ion-concentrations is determined. We find that, in the absence of 
reactions, unequal mobility of ions generate nontrivial features in the 
macroscopically observable concentration profiles. 
Switching on the reaction $A^-+B^+\rightarrow C$ leads to the formation of  
a localized, diffusive reaction front and one finds that
the properties of the front (e.g. effective diffusion constant) are 
affected by the background ions. The consequences of this effect on the 
formation of Liesegang patterns is discussed.
\end{abstract}

\begin{multicols}{2}
\narrowtext

\section{Introduction}
\label{Introduction}
The reaction-diffusion process $A+B\rightarrow C$
has been discussed for a long time. This conceptually 
simple process displays a rich variety of phenomena
(nonclassical reaction kinetics \cite{{ovchzeldo},{Toussaint}}, 
clustering and segregation \cite{{Kangredner},{zumofen}}, 
front formation \cite{{GR},{Leyvredner}}) and, depending on the
interpretation of $A$ and $B$ (particles, quasi-particles, 
topological defects, chemical reagents, etc.), it provides a model 
for a number of phenomena in physics, chemistry, and biology.

In many cases of interest, $A$ and $B$ are ions ($A^-$ and $B^+$) 
and these ions are initially separated from each other. An example 
we shall discuss below is the formation of Liesegang bands 
\cite{{liese},{Henisch}} where an electrolyte $A^-{\hat A}^+$
diffuses into a gel column containing another electrolyte 
${\hat B}^-B^+$. The concentration of $A$-s is taken to be 
much larger than those of the $B$-s, thus the reaction front 
$A^-+B^+\rightarrow C$ moves along the column.
An appropriate choice of reagents then leads to quasiperiodic
precipitation ($C\rightarrow D$) in the wake of the front (Fig.1).


\begin{figure}[htb]
\centerline{
        \epsfysize=0.155\textwidth
        \epsfbox{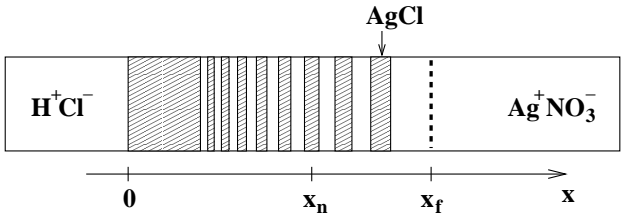}
           }
\vspace{0.2cm}
\caption{Schematic picture of Liesegang phenomena. The correspondence with the 
notation in the text is given by $A^-=Cl^-$, ${\hat A}^+=H^+$ 
(outer electrolyte); $B^+=Ag^+$, ${\hat B}^-=NO_3^-$ (inner electrolyte); and 
$D=AgCl$ (precipitate). The initial interface between 
electrolytes is at $x=0$. The precipitation bands (shaded regions) emerge 
in the wake of the moving reaction-diffusion front (dashed line at $x_f$).}
\vspace{0.2cm}
\label{Fig1}
\end{figure}


In general, the background ions (${\hat A}^+$ and 
${\hat B}^-$) are expected to play a role in a process described above. 
Nevertheless, the usual approach is to neglect them and consider only 
a contact interaction between neutral reagents $A$ and $B$. This 
approximation is based on the argument that the background ions provide 
only screening and, furthermore, the screening length is much smaller 
than the scale of concentration variations relevant in the formation of 
a macroscopic pattern.  
Although the argument sounds compelling, one should note 
that the background ions may generate macroscopic effects 
even if the screening length is negligible. Indeed, if the 
mobility of one of the background ions 
(${\hat A}^+$ in the Liesegang case) is much smaller than 
the other mobilities then the motion and the properties 
of the reaction front are altered. 
Since the properties of the reaction front are crucial in determining 
the pattern \cite{{dee},{ADMRZ},{ADMR}} one expects that the presence of
background ions gives rise to macroscopic changes in the observed 
patterns.  

Our aim with this work is to verify the above expectation and to
investigate how the diffusion and front formation are affected 
by unequal mobilities of background ions. More precisely, 
we shall study the time evolution of ion-concentrations in the process 
\be
A^{^-}+\,{\hat A}^{^+}+\,B^{^+}+\,{\hat B}^{^-}\rightarrow \; 
C\,+\,{\hat A}^{^+}+\,{\hat B}^{^-}
\label{process}
\ee
where the reaction product $C=A^{^-}B^{^+}$ is assumed to
vanish from the system. 
The process starts at $t=0$ form an initial condition where
the electrolytes $A^-{\hat A}^+$ and $B^+{\hat B}^-$ 
are separated and their concentrations ($a^+, {\hat a}^-, b^-, {\hat b}^+$) 
are constant in the left ($x<0$) and right ($x>0$) half-spaces, respectively 
\bea
a^-(x,t=0)=&{\hat a}^+(x,t=0)=a_0&\theta (-x) \nonumber\\
b^+(x,t=0)=&{\hat b}^-(x,t=0)=b_0&\theta (x)
\label{initcond}
\eea
where $\theta(x)$ is the step function.
Such an initial state with $a_0\gg b_0$ 
is actually used in Liesegang experiments, 
and this choice is also motivated by the fact that investigations of 
front formation from such initial state have proved to be 
instrumental in understanding the $A+B\rightarrow C$ process \cite{GR}.

The study of motion of ions is not an easy task and 
we must simplify the problem to make it tractable. 
We believe, however, that our approximations listed below 
are appropriate at least for the description of the Liesegang 
experiments.

\begin{enumerate}

\item It is assumed that the phenomena can be described by 
reaction-diffusion equations. 
This appears to be a correct assumption for reactions taking place in a 
gel where convection is absent. 

\item The screening length is assumed to be negligible and 
screening is taken into account by enforcing local electroneutrality. 
At characteristic ion-concentrations ($10^{-3}\rm{M}-1\rm{M}$)
present in Liesegang experiments, 
the screening length is indeed small ($\sim 10^{-9}m$) compared both to the 
characteristic diffusion length ($\sim 10^{-2}m$) and to the width 
of the reaction zone ($\sim 10^{-6}m$). Further discussion can be
found in Sec.\ref{electroneutral}.

\item The concentration profiles are assumed to depend only on 
one spatial coordinate ($x$ in Fig.1). Although a one dimensional geometry 
can be set up in experiments on Liesegang phenomena (the length of 
the gel column can be made much larger than its width), 
one should note that the 
finite extent of the sample in the transverse direction poses 
nontrivial problems with edge effects. It appears, however, that these effects  
can be neglected since the final pattern is usually one dimensional to a 
good accuracy.

\item The mobility of the 
reagents and of the background ions are, in general, 
different. For simplicity, we shall consider the case 
with one of the background ions having a significantly distinct 
diffusion coefficient
\be
D_{a}=D_{b}=D_{{\hat b}}\equiv D\not=D_{{\hat a}}\equiv {\hat D} \, .
\label{diffconstant}
\ee
This is just a technical assumption to keep the number of parameters 
small and, this also appears to be the most interesting case for 
Liesegang phenomena where $a_0\gg b_0$.

\end{enumerate}
\vspace{0.2truecm}

Once the above approximations are made one arrives at a problem that can be 
studied numerically and, in some limits, analytically. The process is
now simple enough so that the numerical analysis is not hindered by 
computer time and memory problems, or by difficulties arising 
from discretization. 

In order to arrive at the results, we shall proceed as follows. 
First we discuss how to take into account the electroneutrality
constraint in the reaction-diffusion equations (Sec.\ref{electroneutral}). 
Then the case without 
reaction is studied and we show that interesting concentration profiles 
emerge even in the pure diffusion process (Sec.\ref{noreaction}). The
effects of reactions are considered in (Sec.\ref{reactionfront}) where the 
properties of the reaction front are calculated.
Finally, the implications for understanding the Liesegang phenomena is
discussed in Sec.\ref{lieseimpli}.

\section{Equations in the electroneutrality approximation}
\label{electroneutral}
In a medium such as a gel, the ions move by diffusion and, 
in the presence of an electric field ${\vec E}=-\nabla \f$, 
the flux of ions ${\vec j}_i$ is given by
the Nernst-Planck relation \cite{Rubinstein}
\begin{equation}
{\vec j}_i={\vec j}_{i\textrm{,diff}}+{\vec j}_{i\textrm{,drift}}=
-D \left(\nabla n_i + \frac{z_i}{\f_0}n_i\nabla \f \right) .
\label{NernstPlanck}
\end{equation}
Here $n_i$ is the concentration of $i$-th ions of integer charge $z_i$, 
$D_i$ is their diffusion coefficient, and $\varphi_0=RT/F$ is a constant 
combined of the temperature $T$, the gas constant $R$, and the 
Faraday number, $F$. 
The potential $\f$ is determined from the Poisson equation
\be
\Delta \f = -\frac{F}{\e_r \e_0 } \sum_i z_i n_i
\label{Poisson}
\ee
where $\e_0$ is the permittivity of free space while $\e_r$  
is the dielectric constant of the system. 

An important quantity in ionic diffusion is the Debye length $r_D$ that 
gives the characteristic length-scale associated with charge inbalances
\be
r_D=\sqrt{\frac{\e_r\e_0RT}{F^2n_0}} \, ,
\label{r_D}
\ee
where $n_0$ is the characteristic scale of ionic concentrations.
In a Liesegang experiment, one usually has $n_0\approx 
10^{-3}\rm{M}-1\rm{M}$
and the process takes place in an aqueous medium ($\e_r\approx 80$). Thus  
$r_D\approx 10^{-10}m-10^{-8}m$ 
and one can see that $r_D$ is much smaller than the scale   
of the macroscopic pattern (e.g. the 
width of the bands, $\approx 10^{-3}m-10^{-2}m $). As a consequence,
one can use the electroneutrality approximation that consists 
of replacing eq.(\ref{Poisson}) by the constraint 
\be
\sum_i z_i n_i =0 \, .
\label{enconstraint}
\ee
Denoting now the rate of reaction of the $i$-th ion with the others 
by $R_i(\{n\})$ and assuming that the reaction does not violate 
electroneutrality ($\sum_iz_iR_i=0$),   
eq.(\ref{NernstPlanck}) together with the
constraint (\ref{enconstraint}) yields the following equations for the 
time-evolution of the concentration fields 
\be
\p_t n_i= D_i \left[ \Delta n_i -
z_i\nabla\cdot \big( n_i {\vec {\cal E}} \big)\right] - R_i(\{n\})
\label{eqiondiff}
\ee
where the appropriately scaled electric field that arises from the 
electroneutrality constraint is given by
\be
{\vec {\cal E}}=\frac{\sum_i z_i D_i \nabla n_i}{\sum_i z_i^2 D_i n_i} \,  .
\label{elfield}
\ee
It should be noted that there is an extra term in ${\vec {\cal E}}$
if a steady global current flows through the system. Such a current 
is not present in the Liesegang problem and, trying to keep the 
discussion as simple as possible, we shall assume that the global 
current is zero.

For the process of actual interest 
(\ref{process}), reaction takes place only between 
the ions $A^-$ and $B^+$ and their rate of reaction is given
by $ka^-b^+$ where $k$ is the rate constant. Thus the above
equations in a one dimensional geometry take the form 
\bea
\p_t a^-&=& D \Big[ \p^2_x a^- +
\p_x ( a^-  {\cal E} )\Big] - ka^-b^+
\label{1deq1}\\
\p_t b^+&=& D \Big[ \p^2_x b^+ -
\p_x ( b^+ {\cal E} ) \,\Big]- ka^-b^+
\label{1qeq2}\\
\p_t {\hat a}^+&=& {\hat D} \Big[\p^2_x {\hat a}^+ -
\p_x ( {\hat a}^+ {\cal E} )\Big] 
\label{1qeq3}\\
\p_t {\hat b}^-&=& D \Big[ \p^2_x {\hat b}^- +
\p_x ({\hat b}^-  {\cal E} )\,\Big] 
\label{1deq4}
\eea
with 
\be
{\cal E}=\frac{D\p_x(-a^-+b^+-{\hat b}^-)+{\hat D}\p_x{\hat a}^+}
{D(a^-+b^++{\hat b}^-)+{\hat D}{\hat a}^+} \quad .
\label{1dE}
\ee
Equations (\ref{1deq1}-\ref{1dE}) together with the 
initial conditions (\ref{initcond}) provide the mathematical
formulation of our problem. 

Before turning to the solution of the above equations let us mention
that the diffusion-reaction problem of ions 
in one-dimensional geometry can be tackled numerically without assuming 
the electroneutrality condition. The only difficulty is that the 
discretization of space must be on a finer scale than the Debye length and
so, in the range of physical parameters where $r_D$ is exceedingly small,
the calculation becomes impractical. One expects (and we verified it 
for some cases) that the 
solution of the full problem approaches the solution of the 
corresponding "electroneutral" problem as $r_D$ is decreased. 


\section{Concentration profiles with no reactions}
\label{noreaction}

Let us begin the analysis of eqs.(\ref{1deq1}-\ref{1dE}) by considering 
the case of no reactions ($k=0$) and let us further restrict 
our study to the case of $a_0\gg b_0$ corresponding to the 
Liesegang initial conditions. The limit of $b_0=0$ is especially simple 
and treated in textbooks. In this case, the two ions $A^-$ and
${\hat A}^+$ must move together thus an electric field is generated 
that slows down the more mobile ions and accelerates the slower ions.
The result is an effective diffusion with a 
diffusion coefficient $D_{\rm eff}=2D{\hat D}/(D+\hat D)$ \cite{echemtext}.

The presence of a small amount of $B$-s ($a_0\gg b_0$) does not significantly 
change the motion of $A$-s. The ions $A^-$ and
${\hat A}^+$ can separate now but only by an small amount that is compensated 
by the motion of $B$-s. 
On Fig.{\ref{Fig:noreaction1}}, we can see the results for 
the case of $b_0/a_0=0.01$ and $\hat D/D=0.1$ (slow background ions 
$\hat A^+$). 
\begin{figure}[htb]
\centerline{
        \vspace{-0.8cm}
        \epsfysize=0.3\textwidth
        \epsfbox{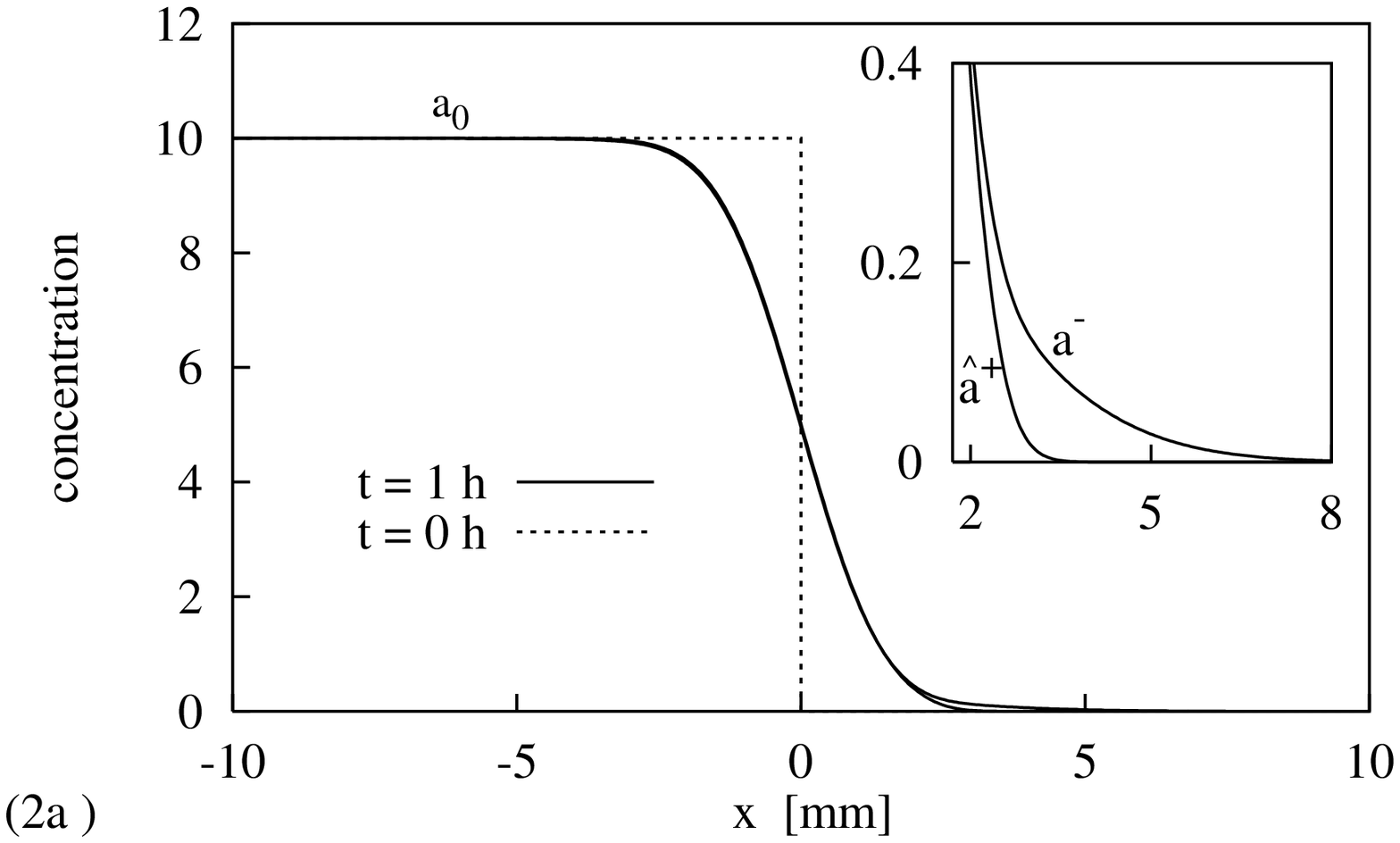}}      
\end{figure}
\begin{figure}[htb]
\centerline{
        \vspace{-0.6cm}
        \hspace{0.15cm}
        \epsfysize=0.295\textwidth
        \epsfbox{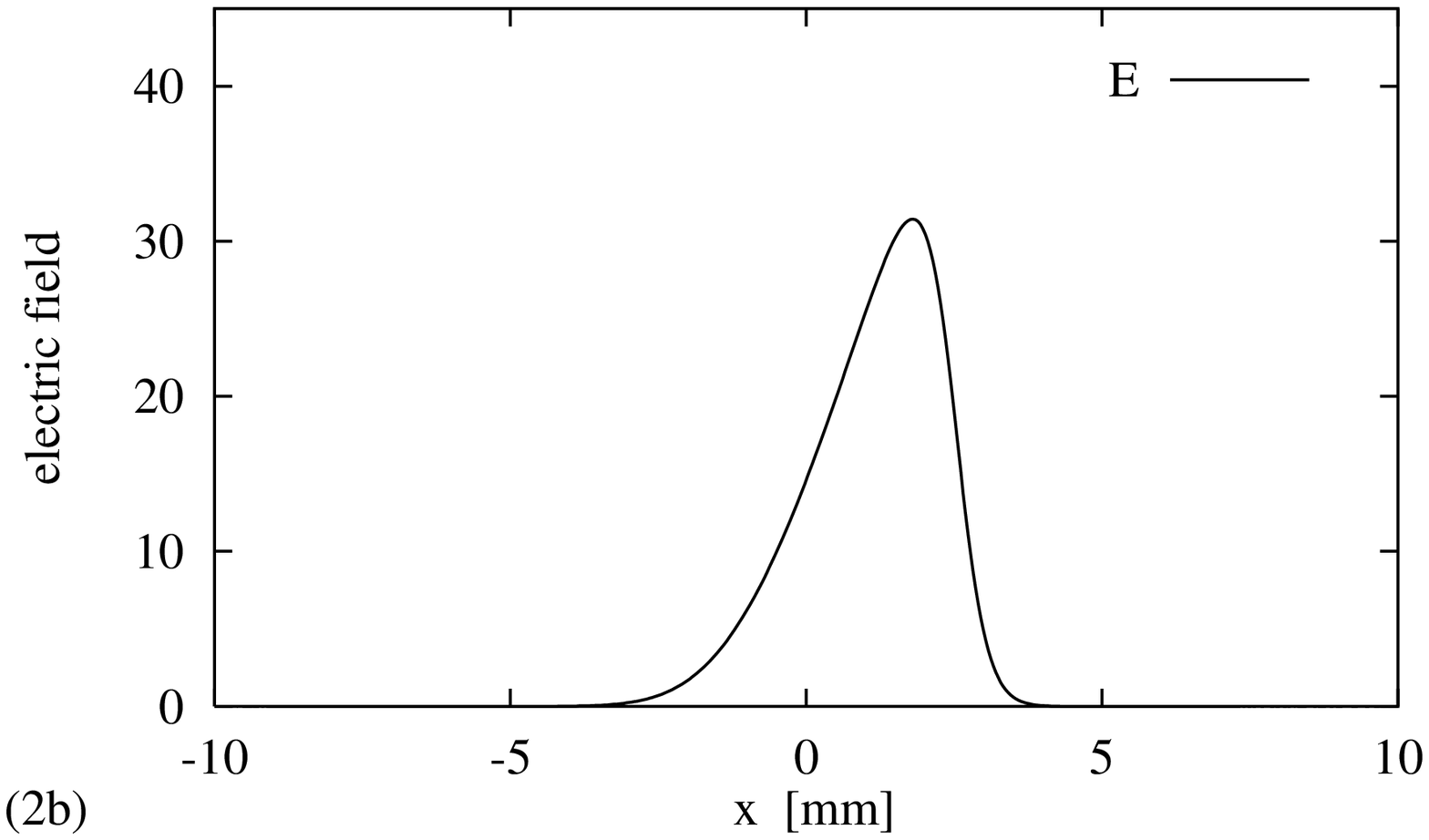}}      
\end{figure}
\begin{figure}[htb]
\centerline{
        \hspace{-0.5cm}
        \epsfysize=0.32\textwidth        
        \epsfbox{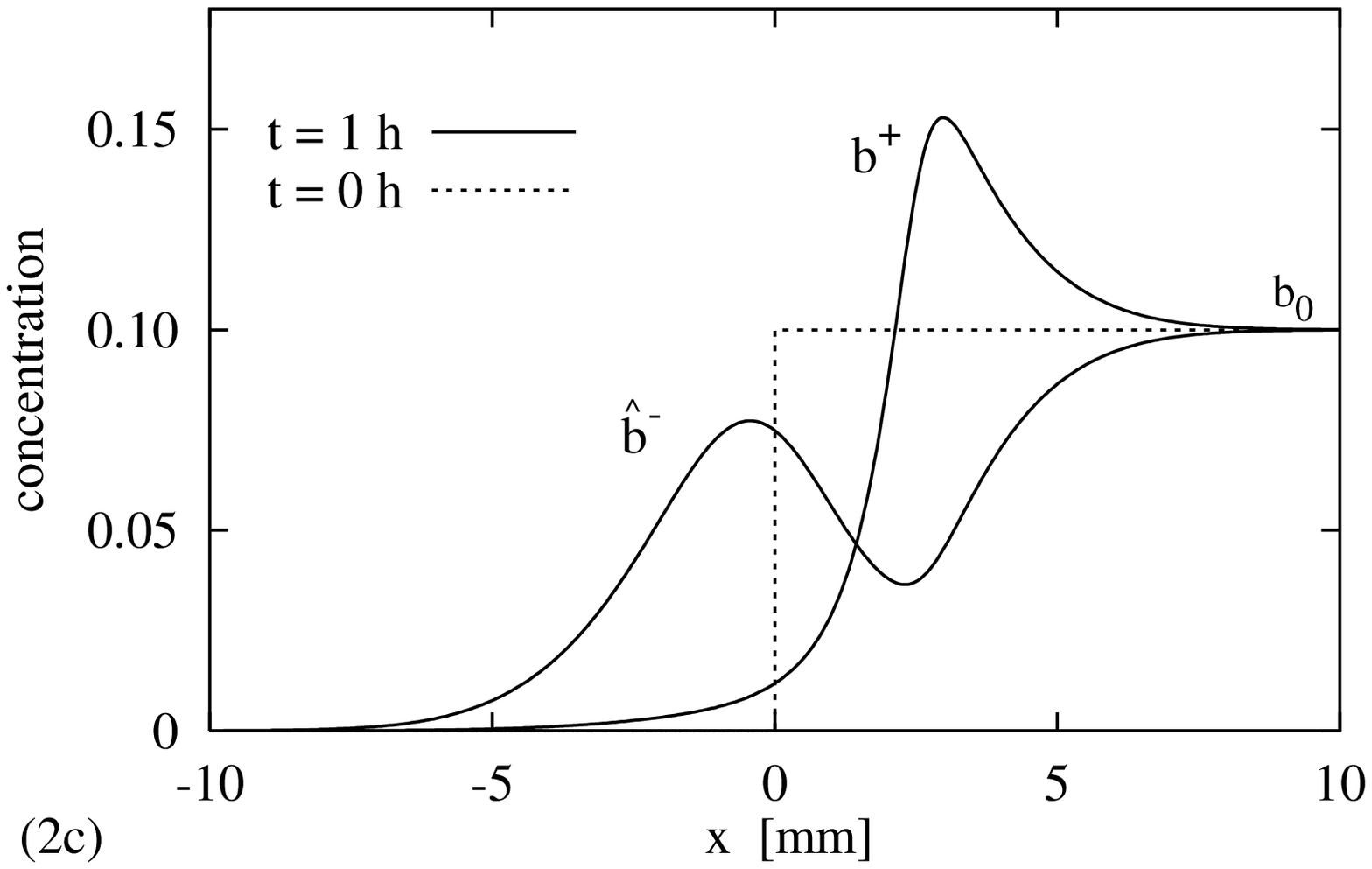}}
\vspace{0.2cm}
\caption{Concentration profiles of the $A$ and $B$ ions 
[(2a) and (2c), respectively) and the electric field (2b) generated 
by them. The concentration is measured in $0.1M$ while the unit of 
the electric field is $V/m$. The results are for the case of $b_0 =0.01a_0$ 
with the diffusion coefficients given by $D=10^{-9}m^2/s$ and $\hat D=0.1D$. 
Inset shows that region where relative separation of the $A$ 
ions is significant.}
\label{Fig:noreaction1}
\end{figure}
The electric field is mainly generated by the motion of the majority $A$ ions 
and, in turn, this field is the determining factor in the motion of the 
minority $B$ ions. 
Since this field, shown in Fig.{\ref{Fig:noreaction1}}b,
moves the $\hat A^+$ ($A^-$) ions in the $+x$ ($-x$) direction, similar 
effect is felt by the $B$ ions. Indeed, as one can see 
in Fig.{\ref{Fig:noreaction1}}c, the $B^+$-s are repelled from the region 
the $A$ ions moved into, while the ions $\hat B^-$ are pulled through 
this region. As a result a region emerges where the ionic and diffusive 
drift of the $B$-s are in opposite directions.

It should be noted that the profiles shown in Fig.2 keep their shape in time.
The pictures at $t^\prime$ are obtained from those at $t=1h$ by 
rescaling the $x$-axis by a factor $\sqrt{t^\prime/t}$. 
This numerical observation is the consequence of the fact that the
Debye length is zero and the initial conditions (\ref{initcond}) do not 
contain any length-scales. 
As a consequence,
all the length-scales are diffusive lengths proportional to $\sqrt{t}$. 
The above argument can be seen to work explictely in
the limit $\hat D=0$ where an analytic calculation \cite{ungerdip} 
gives the concentration profiles that
can be expressed through error functions of argument $x/\sqrt{t}$.

The complexity of concentration profiles shown in Fig.2a and 2c  
suggest that if reactions are switched on between the ions $A^-$ and $B^+$ 
then the emerging reaction front may be rather different from the case 
of neutral reagents. This is what we shall study in Sec.\ref{reactionfront}. 


\section{Reaction front}
\label{reactionfront}

The full reaction-diffusion process is described by 
eqs.(\ref{1deq1}-\ref{1dE}) and the solution of these equations 
with initial condition (\ref{initcond}) provides the 
description of the reaction front. Indeed, once the 
concentration profiles are known, the location and the time evolution of 
the production of $A^-+B^+\rightarrow C$ particles is given by
\be
R(x,t)=ka^-(x,t)b^+(x,t)\, .
\label{reactionrate}
\ee
The properties of $R(x,t)$ are well known for the case of neutral
reagents (${\cal E}=0$) \cite{GR}. In that case, the reaction
takes place in a narrow, moving region whose width is much smaller
than the diffusive scales. The motion of the reaction zone is
`diffusive' characterized by a diffusion constant $D_f$
\be
x_f = \sqrt{2D_ft} \, . 
\label{x_f}
\ee
Another important feature of the front is that it
leaves behind a density of $C$-s \cite{ADMRZ}
\be
c_0=\int_0^\infty R(x,t)dt \,,
\label{c_0}
\ee
that is independent of $x$.

The parameters $D_f$ and $c_0$ can easily be determined for the 
neutral case by exploiting the smallness of the width of the reaction zone. 
The reaction zone is replaced by a point where the diffusion equations
are supplemented by boundary conditions and as a result the parameters
$D_f$ and $c_0$ are given as functions
of $a_0$, $b_0$, $D_a$ and $D_b$ \cite{{GR},{Koza}}.

The presence of a localized, diffusive front is an essential 
ingredient in the theories of Liesegang phenomena \cite{{dee},{ADMRZ},{ADMR}},
and the parameters of the front (especially $D_f$ and $c_0$)
are known to influence the properties of the patterns. 
Thus the next step is now to
find out how the above picture is modified as a result of the ionic
character of the reagents. 

Eqs.(\ref{1deq1}-\ref{1dE}) with initial condition (\ref{initcond})
can be studied by straightforward numerical methods and one finds
that the localized-diffusive-front picture 
does hold and, furthermore, the scaling properties (\ref{x_f}-\ref{c_0}) 
also remain valid when the ionic interactions are switched on. The 
actual values of the parameters $D_f$ and $c_0$, however, are affected by the 
presence of background ions. 

In order to understand how these results arise, let us begin with the 
numerical observation that the reaction front remains narrow even if the
ionic interactions are switched on. Indeed, for characteristic 
values of $a_0\approx 100b_0\approx 1M$,
$D_a\approx D_b\approx 10^{-10}m^2/s$ and $k\approx 10^{10}(Ms)^{-1}$
\cite{Pilling}, we find that the
width is in the mesoscopic range ($\sim 10^{-6}m$) at all times
available in a Liesegang experiment. Thus, on diffusive length-scales, 
the reaction zone can be treated as a point (as in the neutral case) 
and one arrives at equations with no reaction terms
\be
\p_t n_i= D_i \left[ \p^2_x n_i -
z_i\p_x \big( n_i {\cal E} \big)\right]\, .
\label{n_i}
\ee
The reactions are taken into account by the following 
boundary conditions at the front
\bea
a^-(x_f)=b^+(x_f)=0 \nonumber \, ,\\
|j_{a^-}(x_f)|=|j_{b^+}(x_f)| \, .
\label{boundcond}
\eea
The meaning of the above conditions is that the concentrations of the 
reagents are zero at the front
and the flux of ions $A^-$ and $B^+$ to the reaction zone are
equal.

Let us now suppose that $x_f(t)$ and $n_i(x,t)$ are the solutions
of the above equations (\ref{n_i}-\ref{boundcond}) with the initial
condition (\ref{initcond}). Then one can easily verify that the
front-position $\lambda \, x_f(t/\lambda^2)$ and the
concentrations $n_i(x/\lambda,t/\lambda^2)$ also solve the same
problem for an arbitrary $\lambda>0$ (note that the initial conditions 
do not contain any length-scale). 
Thus the functions $n_i(x,t)$ and $x_f(t)$ must satisfy the conditions
$n_i(x,t)=n_i(x/\lambda,t/\lambda^2)$ and $x_f(t)=\lambda \,
x_f(t/\lambda^2)$. As a consequence, we find that
the concetration profiles obey the following scaling form
\be
n_i(x,t)=\Phi_i\left(\frac{x}{\sqrt{t}}\right)
\label{Phi}
\ee
and the front moves diffusively even if the ionic
interactions are taken into account
\be
x_f \sim \sqrt{t}\,.
\label{xf-2}
\ee
The above relationship (\ref{xf-2}) defines the diffusion
constant $D_f$ through $x_f=\sqrt{2 D_f t}$.

The scaling of the concentrations (\ref{Phi}) together with 
equation (\ref{1dE}) imply scaling for the 
electric field 
\be
{\cal E}(x,t)=\frac{1}{\sqrt{t}} \Psi\left(\frac{x}{\sqrt{t}}\right) \,.
\label{Psi}
\ee

These scaling results allow us to investigate the production of the $C$
particles. The number of the $C$-s arising in the reaction zone
in a unit time is given by the flux of one of the reagents (e.g. $j_{a^-}$)
entering the front. According to eqs.(\ref{Phi}-\ref{Psi})
$j_{a^-}$ at $x_f$ is proportional to $1/\sqrt{t}$ and the velocity
of the front decays in time in the same way, $x_f\sim 1/\sqrt{t}$. 
It follows then that the density of the $C$-s emerging in the wake 
of the front is a constant
\be
c=\frac{j_{a^-}}{\dot x_f}=\textrm{const.}=c_0
\label{c0-2}
\ee

\begin{figure}[htb]
\centerline{
        \vspace{-0.8cm}
        \epsfysize=0.33\textwidth
        \epsfbox{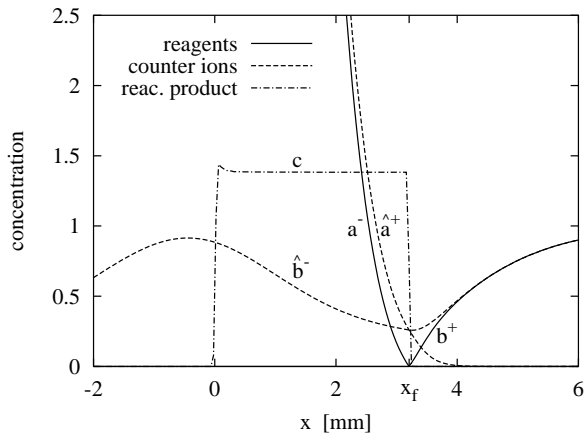}}
\vspace{1.2cm}
\caption{Concentration profiles of the $A$ and $B$ ions when the
reaction is switched on. The initial state is given by
eq.(\ref{initcond}). The results are shown at time $1h$
for the case of $b_0 =0.01a_0=0.01M$ with the diffusion
coefficients given by $D=10^{-9}m^2/s$ and $\hat D=0.1D$.
The concentration is measured in units of $0.01M$.}
\label{Fig:reaction}
\end{figure}

The results (\ref{Phi}-\ref{c0-2}) given by the above analytical
argument have been confirmed by computer simulations. An example of 
such numerical calculation can be seen in 
Fig.\ref{Fig:reaction}.

Having established the same scaling properties of the front 
(\ref{xf-2},\ref{c0-2}) as in the case of neutral reagents,
we turn now to the actual values of the parameters $D_f$ and $c_0$.
Since the motion of the reagents is
modified by the electric field (\ref{Psi}), one expects that 
$D_f$ and $c_0$ will depend not only on the properties of the 
reagents but on the properties of the background ions, as well. 

We studied the effect of the background ions by changing the
diffusion coefficient $\hat D$ (\ref{diffconstant}) and
keeping all the other parameters ($a_0, b_0, D$) fixed.
The numerical results for $D_f$ and $c_0$ 
as functions of $\hat D$, are shown in Fig.\ref{Fig:front}.
As one can see, $c_0$ is does not change significantly in
the physically relevant range of $0.1<\hat D<10$ (Fig.\ref{Fig:front}b).
The reason for this insensitivity of $c_0$ is that 
the density of the reaction product for $a_0\gg b_0$ and $D_a\approx D_b$ 
is mainly determined by the concentration $b_0$ \cite{ADMRZ}. 

\begin{figure}[htb]
\centerline{
        \hspace{-0.8cm}
        \epsfysize=0.33\textwidth
        \epsfbox{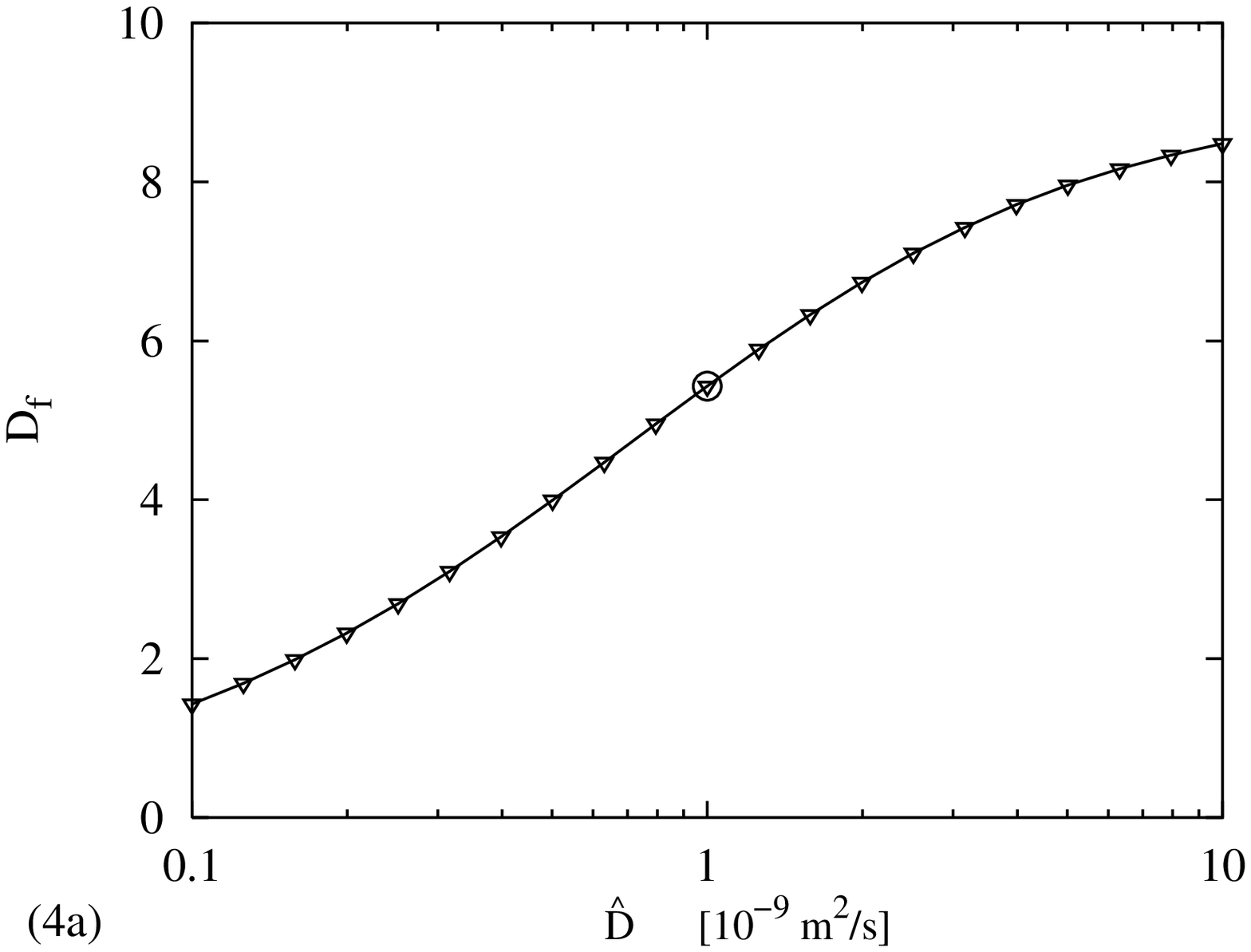}}
\end{figure}
\begin{figure}[htb]
\centerline{
        \hspace{-0.8cm}
        \epsfysize=0.33\textwidth
        \epsfbox{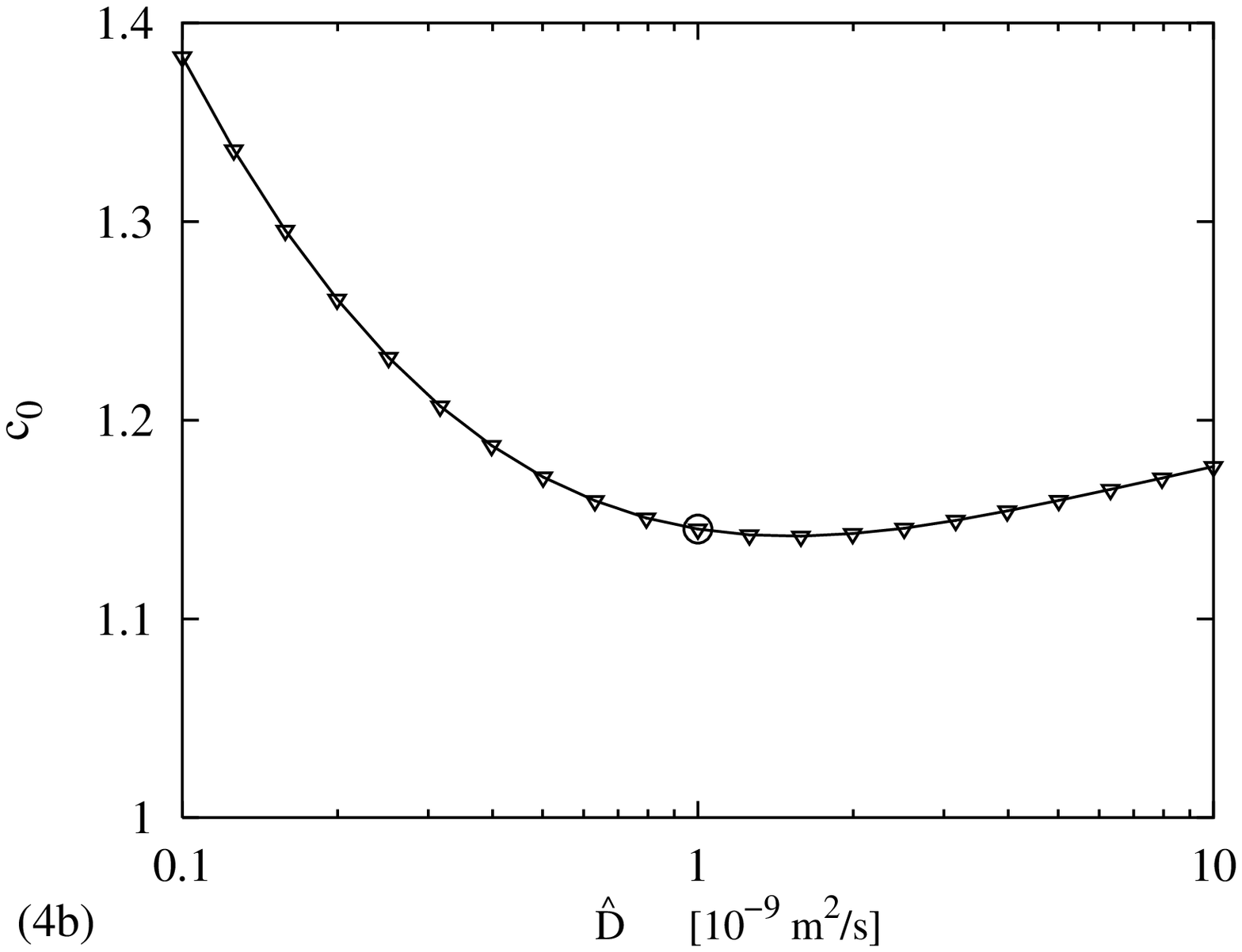}}
\vspace{0.5cm}
\caption{The diffusion coefficient of the reaction front ($D_f$) and
the concentration of the reaction product ($c_0$). $\hat D$ is the
diffusion coefficient of the $\hat{A}^+$ ions while D for the
other ions is $10^{-9}m^2/s$. The units of $D_f$ and
$c_0$ are $10^{-9}m^2/s$ and $0.01M$, respectively. The initial
concentrations are given by $b_0=0.01 a_0=0.01 M$. The points 
indicated by the circles correspond to the case when the background 
ions have no effect on the dynamics.} 
\label{Fig:front}
\end{figure}

The parameter $D_f$ is much more sensitive to the mobility of the
counter ions as shown in Fig.\ref{Fig:front}a. 
Although the motion of the front is 
determined by the interplay of all four types of ions and the process 
is rather complex, the result in Fig.\ref{Fig:front}a can be easily
understood. For $a_0\gg b_0$, the main effect comes from the 
counter ions $\hat{A}^+$ slowing down or speeding up of the motion 
of the $A^-$-s. If the diffusion coefficient $\hat D$
is smaller than $D$, the $A^-$ ions are pulled back by the
$\hat{A}^+$-s (otherwise the slow $\hat{A}^+$ ions would form
positive charge density in the left region) thus fewer $A^-$ particles
enter the front which yields a smaller value of $D_f$. Similar argument 
leads to the opposite effect for the case of $\hat D > D$.
The case ($\hat D = D$) is special in the sense
that the electric field (\ref{1dE}) vanishes and the result 
corresponds to the case of neutral reagents.

In the next section we turn to the theory of Liesegang phenomena
in order to demostrate the relevance of the above results in the 
description of a relatively simple pattern-forming process.


\section{Implications for Liesegang theories}
\label{lieseimpli}

Liesegang patterns described in Sec.\ref{Introduction} 
have been much investigated for about a 
century~\cite{{liese},{Henisch},{bibli}}.  
The gross features of {\it normal} patterns in reproducible 
experiments are rather simple, namely the distance 
between consecutive bands $x_{n+1}-x_n$ increases with band order $n$ 
and the positions of the bands obey   
a {\it spacing law}~\cite{{jabli},{time}}:
\begin{equation}
\frac{x_{n+1}}{x_n} \equiv 1+p_n \stackrel{{\scriptstyle n \gg 
1}}{\longrightarrow} 1+p \quad ,
\end{equation}  
where $1+p$ is called the spacing coefficient and $p>0$. 

Currently, the Liesegang phenomenon is mainly studied 
as a nontrivial example of pattern formation 
in the wake of a moving front \cite{{dee},{luthi}}
and the theories of normal patterns revolve around the 
calculation of $p$. 
The main feature of these theories is that the 
precipitate appears as the system goes through some 
nucleation- \cite{{dee},{luthi},{Ostwald},{Wagner},{Prager},{zeldo}}, 
spinodal- \cite{ADMR}, or coagulation \cite{Chatter}, thresholds.
Most of these theories are rather complicated, however, 
and have been developed only recently to the level \cite{{ADMRZ},{ADMR}} 
that $p$ can be investigated in detail and, in particular, its dependence
on the initial concentrations $a_0$
and $b_0$ can be determined, and connection can be made 
to the experimentally established Matalon-Packter law 
\cite{{Matalon},{Packter}}.

None of the above theories address the question of how the Liesegang 
patterns are affected by the presence of background ions although 
the existence of such an effect is expected. Indeed, let us take, 
for example, the expression for $p$ obtained in a simple version of
the {\it nucleation and growth} theory [see eq.(25) in \cite{ADMRZ}]
\be
p\approx\frac{D_c c^*}{D_f(c_0-c^*)}\quad , 
\label{peq}
\ee
where $D_c$ is the diffusion coefficient of the $C$ particles while $c^*$ 
is the threshold concentration of $C$-s. The meaning of $D_f$ 
(diffusion coefficient of the $C$-s) and $c_0$ 
(the concentration of $C$-s left behind the front) is 
the same as defined in this paper.
As one can see from (\ref{peq}), the spacing coefficient depends both 
on $D_f$ and $c_0$. Thus, on the basis of our results 
(see Fig.\ref{Fig:front}), we expect $p$ to be affected by the 
background ions. 

\begin{figure}[htb]
\centerline{
        \epsfysize=0.33\textwidth
        \epsfbox{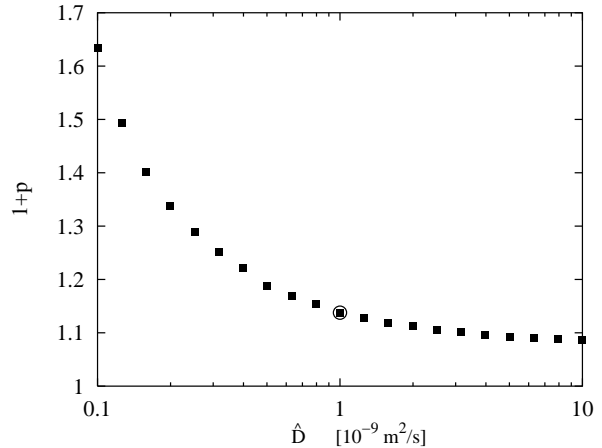}
           }
\vspace{0.7truecm}
\caption{Spacing coefficient as a function of the diffusion coefficient 
($\hat D$) of the background ions ($\hat A^+$) for a case with 
$b_0/a_0=0.01$. 
The circle corresponds to equal diffusion coefficients
where the description in terms of neutral reagents is valid.} 
\vspace{1truecm}
\label{space-coeff.fig}
\end{figure}

In order to put our expectation on a firmer basis, we calculated 
$p$ numerically employing a recent theory \cite{ADMR} where the 
addition of the background ions is straightforward.
The main ingredients in this theory are i) 
a moving reaction front that leaves behind the particles $C$, and ii) a
Cahn-Hilliard type phase-separation dynamics for the $C$ particles. 
This theory yields the spacing law, and the results for $p$ 
are in agreement with the Matalon-Packter law. Thus it appears to 
be a good candidate for the description of the Liesegang process.
Since the reaction front enters the description only as a source in the
Cahn-Hilliard equation [see eq.(3) in \cite{ADMR}], one can study 
the effect of background ions by modifying the source according to 
what has been described in Sec.\ref{reactionfront}. The results of 
our numerical work for a particular case with $b_0/a_0=0.01$ 
(the parameters in the Cahn-Hilliard equation were set to unity) 
is displayed in Fig.\ref{space-coeff.fig} 

As can be seen from Fig.\ref{space-coeff.fig}, $p$ does depend on 
$\hat D$ and, actually, $p$ can change by a factor five 
compared to the neutral case ($\hat D\approx D$) provided 
$\hat D$ decreases by a factor ten. One can also observe that the 
ionic effect is larger when the counterion $A^+$ is slower than $A^-$. 
These obsevations and the overall picture is in agreement with the 
result (\ref{peq}) obtained in the {\it nucleation and growth} theory. 
Indeed, $c_0$ is weakly dependent on $\hat D$ thus the main effect comes 
from $D_f$. As Fig.\ref{Fig:front} shows, $D_f$ is a smooth, monotonically 
increasing function of $\hat D$ and this translates through 
eq.(\ref{peq}) into a monotonically decreasing $p(\hat D)$.

We have thus shown that the backgroung ions cannot be neglected in the 
description of the Liesegang phenomena unless the diffusivities of the
ions are roughly equal. Although this conclusion appears to complicate 
the description significantly, the reassuring aspect of the result is that
all the complications can be absorbed into the parameters  
($D_f$ and $c_0$) of the front. As a consequence, previous ideas about 
the pattern formation remain intact apart from the need of taking 
account of the renormalization of the parameters $D_f$ and $c_0$.

\section{Final remarks}
A general conclusion we can draw from the present work is that 
the dynamics of reaction fronts is strongly altered
if the diffusivities of the reacting ions differ significantly from
those of the  background ions. This conclusion is based on 
the nontrivial density profiles found in a study of the 
the simplest reaction scheme $A+B\rightarrow C$ and assuming 
negligible screening length (electroneutrality approximation). 
We believe, however, that some aspects of our results
(the reaction front can still be characterized by 
effective diffusion constant and it still leaves behind a constant 
density of reaction product) are robust since they appear to 
follow from more general considerations and thus they should be 
applicable to the more complicated cases.

\section*{Acknowledgments}
We thank  M. Droz, P. Hantz, L. Szil\'agyi, and M. Zr\'\i nyi for 
useful discussions. This work has been supported
by Hungarian Research Funds (Grants OTKA T 029792 and FKFP-0128/1997).


\end{multicols}

\begin{references}
\bibitem{time} In addition to the spacing law, 
               there is also a time-law and width law. The former
               states that the time of appearance
               $t_n$ of the $n$ band is related to the position of the 
               band by $x_n\sim \sqrt{t_n}$ while the latter  
               relates the width and the position of the bands 
               ($w_n\sim x_n$).
\end{references}
\end{document}